\documentclass[aps,prd,nofootinbib]{revtex4}
\usepackage{graphicx}
\usepackage{amsfonts}
\usepackage{amsmath}
\usepackage{amssymb}

\begin{document}

\title{Galactic microlensing by Lobo-Parsaei-Riazi phantom wormhole: Paczy\'{n}ski light curves and probabilistic features}
\author{G.F. Akhtaryanova}
\email{akht\_gul@mail.ru}
\affiliation{Zel'dovich International Center for Astrophysics, Bashkir State Pedagogical University, 3A, October Revolution Street, Ufa 450077, RB, Russia}
\author{R.Kh. Karimov}
\email{karimov_ramis_92@mail.ru}
\affiliation{Zel'dovich International Center for Astrophysics, Bashkir State Pedagogical University, 3A, October Revolution Street, Ufa 450077, RB, Russia}
\author{R.N. Izmailov}
\email{izmailov.ramil@gmail.com}
\affiliation{Zel'dovich International Center for Astrophysics, Bashkir State Pedagogical University, 3A, October Revolution Street, Ufa 450077, RB, Russia}
\author{U.K. Khidirov}
\email{umurzokk@mail.com.ru}
\affiliation{Zel'dovich International Center for Astrophysics, Bashkir State Pedagogical University, 3A, October Revolution Street, Ufa 450077, RB, Russia}

\date{19 June 2025}

\begin{abstract}
Gravitational microlensing can provide a possible observational method for distinguishing between the signatures of massive and massless phantom wormholes. In this work, we consider Galactic microlensing by the bounded Lobo-Parsaei-Riazi phantom wormhole (LPR), assuming source stars located in the Galactic Bulge and in the Large Magellanic Cloud (LMC). We derive the weak-field deflection angle up to fourth post-Newtonian order and compute the Einstein radius, Einstein-radius crossing time, and idealized point-source Paczy\'{n}ski-type light curves. We also estimate the optical depth and event rate in a simplified model in which the wormhole lenses are assumed to be gravitationally bound to the Galaxy. The LPR parameter $\gamma$, which is related to the radial equation-of-state parameter by $\omega = 1/ \gamma$, affects the ADM mass, the Einstein radius, and the microlensing timescale. In the massive phantom branch $-1 < \gamma < 0$, the leading deflection term is proportional to $1/r$, and the resulting point-source light curves are Paczy\'{n}ski-like. The massless comparison case $\gamma = 1$ is qualitatively different because the leading $1/r$ term vanishes and gutters may appear. The idealized observational predictions are compared with those for a Schwarzschild black hole.
\end{abstract}

%\pacs{}
\maketitle

%%%%%%%%%%%%%%%%%%  DATE  %%%%%%%%%%%%%%%%%%%

%%%%%%%%%%%%%%%%%%%%%%%%%%%%%%%%%%%%%%%%%
\section{Introduction}
\label{sec1}
%%%%%%%%%%%%%%%%%%%%%%%%%%%%%%%%%%%%%%%%%
We study an application of natural wormholes as gravitational lenses that are threaded by the ubiquitous phantom matter believed to be responsible for late-time cosmic acceleration. The phantom wormhole solution considered below is supported by an anisotropic stress-energy tensor. Its radial pressure obeys $p_r = \omega\rho$, while the tangential pressure is determined independently by the remaining Einstein equations. Therefore, although the source violates the usual energy conditions and is often called "phantom", the parameter $\omega$ should not be identified with the equation-of-state parameter of homogeneous cosmological dark energy. In this work we adopt a phenomenological microlensing setup in which such compact wormhole objects are assumed to be gravitationally bound to the Galaxy and to act as lenses of source stars in the Galactic Bulge and in the Large Magellanic Cloud.

Gravitational microlensing is the effect of light bending by very small
angle in the gravitational field of the lens, in which one or more images of
a light source may appear, and in a special case, an Einstein ring may
appear. Here we assume the lens to be the Lobo-Parsaei-Riazi (LPR) \cite{Lobo:2013} phantom wormhole sourced by exotic matter, that is, matter violating one or
more energy conditions. Such matter causes the spacetime to have a
nontrivial wormhole topology connecting two distant regions of space or even
two universes.

Wormhole spacetimes are valid solutions to theories of gravity that have not
yet been ruled out by experiments and are included in the search for massive
compact halo objects (MACHOs) in the halo region of galaxies \cite{Cramer:1995}. The
concept of wormholes was formalized by Morris and Thorne in their seminal
paper \cite{Morris:1988}. Black holes and wormholes have different topologies, and although
the formation of black holes is understood as a possible end product of the
collapse of matter under appropriate conditions or even as a result of the
collapse of scalar field threaded wormholes \cite{Bronnikov:2011}, the mechanism of the
formation of a classical wormhole itself is still not completely understood.
At the same time, the mouth of a wormhole can simulate the behavior of the
event horizon of a black hole. Features of the observational characteristics
of wormholes include characteristics of accretion disks \cite{Sokoliuk:2022, Yusupova:2025, Karimov:2020, Yusupova:2021}, models of
wormholes with a thin shell, the outer region of which is described by
solutions for black holes \cite{Khaybullina:2014, Akhtaryanova:2025}, characteristics of gravitational lensing
in both weak \cite{Izmailov:2020a, Liu:2022, Ahmed:2025, Lukmanova:2018, Izmailov:2020b, Bronnikov:2019, Kuhfittig:2014, Asada:2017, Eiroa:2001, Tsukamoto:2012, Jusufi:2018a, Jusufi:2018b, Li:2020, Nakajima:2012} and strong field limits \cite{Godani:2021, Izmailov:2019, Sarkar:2026, Nandi:2018} and wormhole shadows \cite{Rahaman:2021, Bugaev:2021, Kumar:2024, Tsukamoto:2021, Ishkaeva:2023, Bronnikov:2025}. Apart from these, there could be other distinctive properties of
wormholes. For instance, wormholes resulting from the merger of two
Schwarzschild black holes can emit gravitational waves \cite{Cardoso:2016}. Such ring-down
emissions can occur also from the perturbations of massive Ellis-Bronnikov
wormholes \cite{Nandi:2017}.

LPR phantom wormhole \cite{Lobo:2013} is asymptotically flat and its local signatures
have been analyzed in the literature from various perspectives \cite{Lukmanova:2016a, Nandi:2016}.
Recently, the weak field deflection of light up to second order by LPR
wormholes with a bounded/unbounded mass function, and with a vanishing
redshift function were studied in \cite{Ovgun:2019}. The authors showed that the
particular choice of the shape function and mass function play a crucial
role in the final expression for the deflection angle of light.

In the Galactic microlensing scenario, phantom wormholes could play
the role of re-defined lenses that take into account the effect of phantom
matter distribution of the background spacetime which, in the present paper,
is the LPR wormhole with bounded mass function. Microlensing properties of
massless Ellis-Bronnikov wormholes were studied in the seminal paper by
Abe \cite{Abe:2010}. Microlensing effects of wormholes were studied in \cite{Safonova:2001, Lukmanova:2016b, Akhtaryanova:2024a, Gao:2023, Gao:2024, Tsukamoto:2018, Tsukamoto:2017, Akhtaryanova:2024b, Toki:2011, Paczynski:1986, Liu:2023}.
Finite-source microlensing and ray-traced light curves have also been investigated for other nonsingular compact objects. Boos and Hu \cite{Boos:2026} combined a fourth-order PPN treatment for pointlike sources with a ray-tracing calculation for extended sources. They showed that the ray-traced light curves approach the point-source prediction as the source radius is reduced, while a finite source size can modify the resulting magnification. Although their compact-object models differ from the LPR wormhole considered here, their analysis provides a useful methodological benchmark and motivates a finite-source ray-tracing extension of the present calculation. The effect of phantom matter is expected to be manifest in the lensing
signatures of the background geometry, particularly in the light curves \cite{Bugaev:2025}. This exciting possibility motivated us to further explore the precise
effects of phantom matter on the microlensing observables in an
astrophysically natural setting.

The purpose of the present paper is to consider the special case of
LPR phantom wormholes and comprehensively study Galactic microlensing by
these wormholes, when the source stars are in the Galactic Bulge or in the
Large Magellanic Cloud. We shall analyze the probabilistic features such as
optical depth and event rate assuming the wormhole lens to be bound to our
Galaxy. We shall analyze the behavior of Paczy\'{n}ski light curves that
typically characterize the distinguishing properties between black holes and
wormholes.

The paper is organized as follows. In Sec. II we shall briefly outline LPR phantom wormholes with bounded mass function. Sec. III works out the light deflection angle correctly up to 4$^{th}$ PPN order and Sec. IV works out Paczy\'{n}ski light curves. In Sec. V, we deal with the probabilistic characteristics of microlensing and conclude in Sec. VI. We choose units such that $G=1$, $c=1$.

%%%%%%%%%%%%%%%%%%%%%%%%%%%%%%%%%%%%%%%%%
\section{LPR phantom wormhole}
\label{sec2}
%%%%%%%%%%%%%%%%%%%%%%%%%%%%%%%%%%%%%%%%%
The general static and spherically symmetric line element representing a phantom wormhole is given by \cite{Lobo:2013} 
\begin{equation}
ds^{2} = -U(r) dt^{2} + \left[ 1 - \frac{b(r)}{r} \right]^{-1} dr^{2} + r^{2} \left(d\theta ^{2} + sin^{2}\theta d\phi^{2}\right),
\end{equation}%
with the redshift function $U(r)=k\left( 1+\frac{\gamma r_{0}}{r}\right) ^{1-1/\gamma},$ where $k$ is a constant of integration.

The shape function takes the form 
\begin{equation}
b(r) = r_{0} + \gamma r_{0}\left[ \left( \frac{r}{r_{0}}\right) ^{\alpha}-1%
\right] ,
\end{equation}%
where $r_{0}$ is the throat radius, $\gamma$, $\alpha$ are dimensionless
constants. In order to satisfy the flaring-out condition at the throat, $%
b^{\prime }(r_{0})<1$, one deduces the additional constraint $\alpha \gamma<1$.
In summary, one has the following restrictions for the parameters 
\begin{equation}
\alpha <1,\ 0<\alpha \gamma<1.
\end{equation}

The equation of state $p_{r}$ $=\omega \rho $, where $p_{r}$ is the radial
pressure and $\rho $ is the source matter density, $\omega $ is a constant, imposes at the throat, the following condition \cite{Lobo:2013} 
\begin{equation}
\alpha \omega \gamma=-1.
\end{equation}%
For the bounded branch we set $\alpha=-1$. Then Eq. (4) gives $\omega=1/\gamma$. Thus the phantom radial equation of state, $\omega<-1$, corresponds to $-1<\gamma<0$. In the LPR notation the finite mass function is defined by
$$b(r)=r_0+2m(r),$$ 
so that, for $\alpha=-1$,
$$b(r)=r_0(1-\gamma)+\frac{\gamma r_0^2}{r},\qquad m(r)=\frac{\gamma r_0}{2}\left(\frac{r_0}{r}-1\right).$$
This $m(r)$ should not be confused with the Misner-Sharp mass $M_{\rm MS}(r) = b(r)/2$.

The metric in this case is given by 
\begin{equation}
ds^{2} = -\left( 1+\frac{\gamma r_{0}}{r}\right) ^{1-\frac{1}{\gamma}}dt^{2}+\left[ 1-%
\frac{r_{0}}{r}\left( 1-\gamma +\frac{\gamma r_{0}}{r}\right) \right] ^{-1}dr^{2}+r^{2}d%
\Omega ^{2},
\end{equation}%
where it is transparent that the geometry is asymptotically flat.

The asymptotic ADM mass is defined by
\begin{equation}
M_{ADM}=\frac{1}{16\pi }\int \int_{S}\overset{3}{\underset{i,j=1}{\sum }}%
\left( \partial _{j}g_{ij}-\partial _{i}g_{ii}\right) n^{i}dS,
\end{equation}
where $S$ is a 2-surface enclosing the active region and $n^{i}$ denotes the
unit outward normal. For the metric (5), we get 
\begin{equation}
M_{\text{ADM}} = \frac{(1-\gamma)r_{0}}{2}.
\end{equation}%
The value $\gamma=1$ gives $M_{\rm ADM}=0$ and corresponds to the massless LPR, or Ellis-type \cite{Ellis:1973, Bronnikov:1973}, limit. This value is outside the phantom branch $-1<\gamma<0$ used for massive LPR lenses and is included only as a comparison case.

The stress-energy tensor of the LPR solution is anisotropic. The radial pressure satisfies $p_r=\omega\rho$, but the tangential pressure $p_t$ is not assumed to obey the same equation of state and, in general, $p_t\ne p_r$. This anisotropy is essential: static twice-asymptotically-flat wormholes are excluded for isotropic fluid sources under the assumptions of the no-go theorem of Bronnikov, Baleevskikh and Skvortsova, whereas anisotropic sources avoid that restriction \cite{Bronnikov:2017}.

In the next section, the deflection angle up to 4th order will be obtained using a modified method Keeton-Petters \cite{Keeton:2005}.

%%%%%%%%%%%%%%%%%%%%%%%%%%%%%%%%%%%%%%%%%
\section{Light deflection angle}
\label{sec3}
%%%%%%%%%%%%%%%%%%%%%%%%%%%%%%%%%%%%%%%%%
We employ Keeton-Petters method \cite{Keeton:2005} that was modified in \cite{Akhtaryanova:2024b} to calculate
the light deflection angle up to 4th order. We start with the general form
of metric in standard coordinates which is given by
\begin{equation}
ds^{2} = -A(r) dt^{2} + B(r) dr^{2} + r^{2} \left( d\theta^{2} + \sin^{2}\theta d\varphi^{2} \right).
\end{equation}
In the weak-field region $r\gg r_0$, we expand the metric functions in powers of the dimensionless quantity $r_{0}/r$ as
\begin{eqnarray}
A(r) &=& 1 - 2a_{1} \left(\frac{r_{0}}{r}\right) 
+ 2a_{2} \left(\frac{r_{0}}{r}\right)^{2} 
- 2a_{3} \left(\frac{r_{0}}{r}\right)^{3} 
+ 2a_{4} \left(\frac{r_{0}}{r}\right)^{4} + ...\; , \\
B(r) &=& 1 + 2b_{1} \left(\frac{r_{0}}{r}\right)
+ 4b_{2} \left(\frac{r_{0}}{r}\right)^{2} 
+ 8b_{3} \left(\frac{r_{0}}{r}\right)^{3}
+ 16b_{4}\left(\frac{r_{0}}{r}\right)^{4} + ...\; ,
\end{eqnarray}
where $a_{1}$, $a_{2}$, $a_{3}$, $a_{4}$, $b_{1}$, $b_{2}$, $b_{3}$ and $b_{4}$ are the coefficients of the PPN expansion. Then the deflection angle up to 4th PPN order is given by
\begin{equation}
\widehat{a}(b_{\rm imp}) = A_{1} \left(\frac{r_{0}}{b_{\rm imp}}\right) + A_{2} \left(\frac{r_{0}}{b_{\rm imp}
} \right)^{2} + A_{3}\left(\frac{r_{0}}{b_{\rm imp}}\right)^{3} + A_{4}\left(\frac{r_{0}
}{b_{\rm imp}}\right)^{4} + ...\; ,
\end{equation}%
where $b_{\rm imp}$ is the impact parameter and
\begin{eqnarray}
A_{1} &=& 2\left(a_{1} +b_{1}\right), \\
A_{2} &=& \pi \left(2a_{1}^{2} + a_{1}b_{1} - a_{2} - \frac{b_{1}^{2}}{4} + b_{2}\right), \\
A_{3} &=& \frac{2}{3} \left(
35a_{1}^{3} + 15a_{1}^{2}b_{1} - 30a_{1}a_{2} - 3a_{1}b_{1}^{2} + 12a_{1}b_{2} - 6a_{2}b_{1} + 6a_{3} + b_{1}^{3} - 4b_{1}b_{2} + 8b_{3}\right), \\
A_{4} &=& \pi \left[30a_{1}^{4} + 12a_{1}^{3}b_{1} -\frac{9}{4}a_{1}^{2}\left(
16a_{2}+b_{1}^{2}-4b_{2}\right) + a_{1} \left( 9a_{3} - 9a_{2}b_{1} + \frac{3b_{1}^{3}}{4} - 3b_{1}b_{2} + 6b_{3}\right)\right. \nonumber \\
&&\left. +\frac{3}{64} \left(96a_{2}^{2} + 16a_{2} \left(b_{1}^{2} - 4b_{2}\right) + 32a_{3}b_{1} - 32a_{4} - 5b_{1}^{4} + 24b_{1}^{2}b_{2} - 32b_{1}b_{3} - 16b_{2}^{2} + 64b_{4}\right)\right].
\end{eqnarray}
The expansion of metric functions in (9)-(10) in powers of $r_{0}/r$ has the form
\begin{eqnarray}
A(r) &=& 1- \left(1 - \gamma\right) \left(\frac{r_{0}}{r}\right) 
+ \frac{\left(1 - \gamma\right)}{2} \left(\frac{r_{0}}{r}\right)^{2} -
\frac{\left(1-\gamma^{2}\right)}{6} \left(\frac{r_{0}}{r}\right)^{3} + \frac{\left(1 + 2\gamma - \gamma^{2} - 2\gamma^{3}\right)}{24} \left(\frac{r_{0}}{r}\right)^{4} + ...\; , \\
B(r) &=& 1 + \left(1 - \gamma\right) \left(\frac{r_{0}}{r}\right) 
+ \left( 1-\gamma + \gamma^{2}\right)
\left(\frac{r_{0}}{r}\right)^{2} 
+ \left(1 - \gamma + \gamma^{2} - \gamma^{3}\right)
\left(\frac{r_{0}}{r}\right)^{3}
+ \left(1 - \gamma + \gamma^{2} - \gamma^{3} + \gamma^{4} \right) \left(\frac{r_{0}}{r} \right)^{4} + ...\; .
\end{eqnarray}
So we can obtain the PPN coefficients
\begin{eqnarray}
a_{1} &=& \frac{\left(1-\gamma\right)}{2}; \quad
a_{2}=\frac{\left(1-\gamma\right)}{4}; \quad
a_{3}=\frac{\left(1-\gamma^{2}\right)}{12}; \quad
a_{4}=\frac{\left( 1+2\gamma- \gamma^{2}- 2\gamma^{3}\right)}{48}, \\
b_{1} &=& \frac{\left( 1-\gamma\right)}{2}; \quad
b_{2}=\frac{\left(1-\gamma+\gamma^{2}\right)}{4}; \quad
b_{3}=\frac{\left( 1-\gamma+\gamma^{2}-\gamma^{3}\right)}{8}; \quad
b_{4}=\frac{\left( 1-\gamma+\gamma^{2}-\gamma^{3}+\gamma^{4}\right) }{16},
\end{eqnarray}
which yield
\begin{eqnarray}
A_{1} &=& 2 \left(1 - \gamma\right), \\
A_{2} &=& \frac{\pi \left(11 - 22\gamma + 15 \gamma^{2}\right)}{16}, \\
A_{3} &=& \frac{ 8 \left(1 - 3\gamma + 4\gamma^{2} - 2\gamma^{3}\right)}{3}, \\
A_{4} &=& \frac{\pi \left( 1145 - 4556\gamma + 8902\gamma^2  - 8812\gamma^3 + 3465\gamma^4 \right) }{1024}.
\end{eqnarray}
The bending angle can then be expressed as
\begin{eqnarray}
\widehat{a}_{LPR}(b_{\rm imp}) &=&2\left( 1-\gamma\right) \left( \frac{r_{0}}{b_{\rm imp}}\right) +%
\frac{\pi \left( 15\gamma^{2}-22\gamma+11\right) }{16}\left( \frac{r_{0}}{b_{\rm imp}}\right)
^{2}-\frac{8\left( 2\gamma^{3}-4\gamma^{2}+3\gamma-1\right) }{3}\left( \frac{r_{0}}{b_{\rm imp}}%
\right) ^{3} \nonumber \\
&&+\frac{\pi \left( 3465\gamma^4-8812\gamma^3+8902\gamma^2-4556\gamma +1145\right) }{1024}%
\left( \frac{r_{0}}{b_{\rm imp}}\right) ^{4} + ...\; . 
\end{eqnarray}

\begin{figure}
	\centering \includegraphics[width=.5\textwidth]{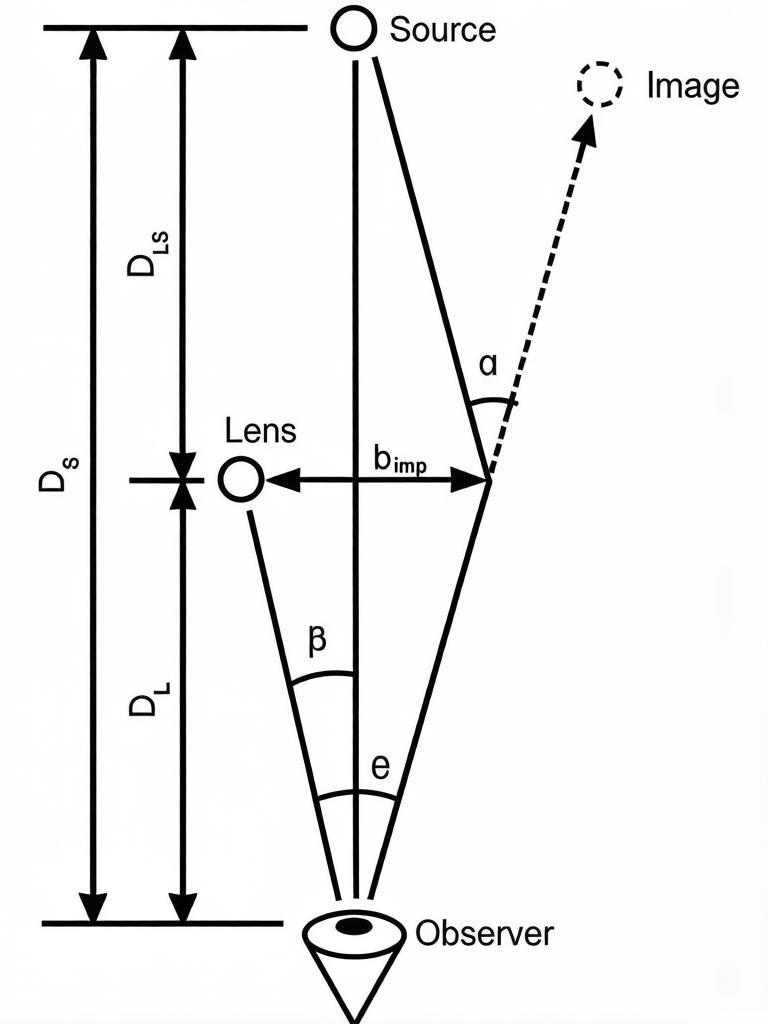}
	\caption{Lens geometry, where $D_{L}$ is the distance from the observer to the lens, $D_{S}$ is the distance from the observer to the source, $D_{LS}$ is the distance from the lens to the source, $b_{\rm imp}$ is the impact parameter of the light, $\protect\beta$ is the angle between lens and source, $\protect\theta $ is the angle between the image and lens, and $\protect\alpha$ is the bending angle between the image and source. The physical Einstein radius $R_E$ is the transverse distance from the lens position to the Einstein ring in the lens plane when $\beta = 0$, and it is related to the angular Einstein radius by $R_{E} = D_{L} \theta_E$. Figure taken from \cite{Abe:2010}.}
	\label{FIG:1}
\end{figure}

%%%%%%%%%%%%%%%%%%%%%%%%%%%%%%%%%%%%%%%%%
\section{Microlensing properties of phantom wormholes}
\label{sec4}
%%%%%%%%%%%%%%%%%%%%%%%%%%%%%%%%%%%%%%%%%
We now turn to the microlensing properties of the LPR wormhole. We emphasize that Eq. (24) is an expansion in the impact parameter $b_{\rm imp}$. If the deflection angle is instead written in terms of the closest-approach coordinate $r$, the relation $b_{\rm imp}=r/\sqrt{A(r)}$ must be used consistently. To second order this gives
\begin{equation}
b_{\rm imp}\simeq r\left[1+\frac{(1-\gamma)r_0}{2r} +\frac{(1-4\gamma+3\gamma^2) r_0^2}{8r^2}\right], 
\end{equation}
and hence the deflection angle up to second order in terms $r_{0}/r$ gives
\begin{equation}
\widehat{a}(r) =
2(1-\gamma)\frac{r_0}{r} + \frac{\pi (15\gamma^2 - 22\gamma + 11) - 16 (1-\gamma)^2}{16} \left(\frac{r_0}{r}\right)^2 +O\!\left( \frac{r_{0}}{r}\right)^{3}.
\end{equation}%
Eqs. (25) and (26) are auxiliary conversion equations between $r$ and $b_{\rm imp}$. The closest-approach coordinate $r$ must not be identified with the lens-plane impact radius $D_{L} |\theta|$. It is clear that when $\gamma=1$ the leading order of the bending angle will be zero.

The source angle $\beta$ between the optical axis (the line joining the wormhole lens and the observer) and the line joining the observer and the source star can be written from the elementary lensing geometry \cite{Abe:2010, Akhtaryanova:2024b} for light traversing the lens on two opposite sides as (see Fig.1 for lensing geometry):
\begin{equation}
\beta = \frac{b_{\rm imp} (r)}{D_{L}} - \frac{D_{LS}}{D_{S}} \widehat{a}(r),
\end{equation}%
where the Euclidean distances from the observer to the lens are $D_{L}$ and to the source is $D_{S}$, the distance from the lens to the source is $D_{LS}=D_{S}-D_{L}$, respectively.

The lens is placed at the origin of the lens plane. Since the lens is spherically symmetric, the light deflection depends only on the positive closest-approach distance $r$, or equivalently on the positive impact parameter $b_{\rm imp}$. However, the image position on the lens plane must be treated as a signed quantity. We denote the signed angular position of an image by $\theta$, or equivalently the signed lens-plane coordinate by $\xi = D_{L} \theta$ as shown in Fig.2. Positive and negative values of $\theta$ do not mean positive or negative radial distance. They simply indicate that the image is formed on one side or the other side of the optical axis. Thus $r = D_{L} |\theta|$, while $\theta$ itself may be positive or negative.

\begin{figure}
	\centering
    \includegraphics[width=.6\textwidth]{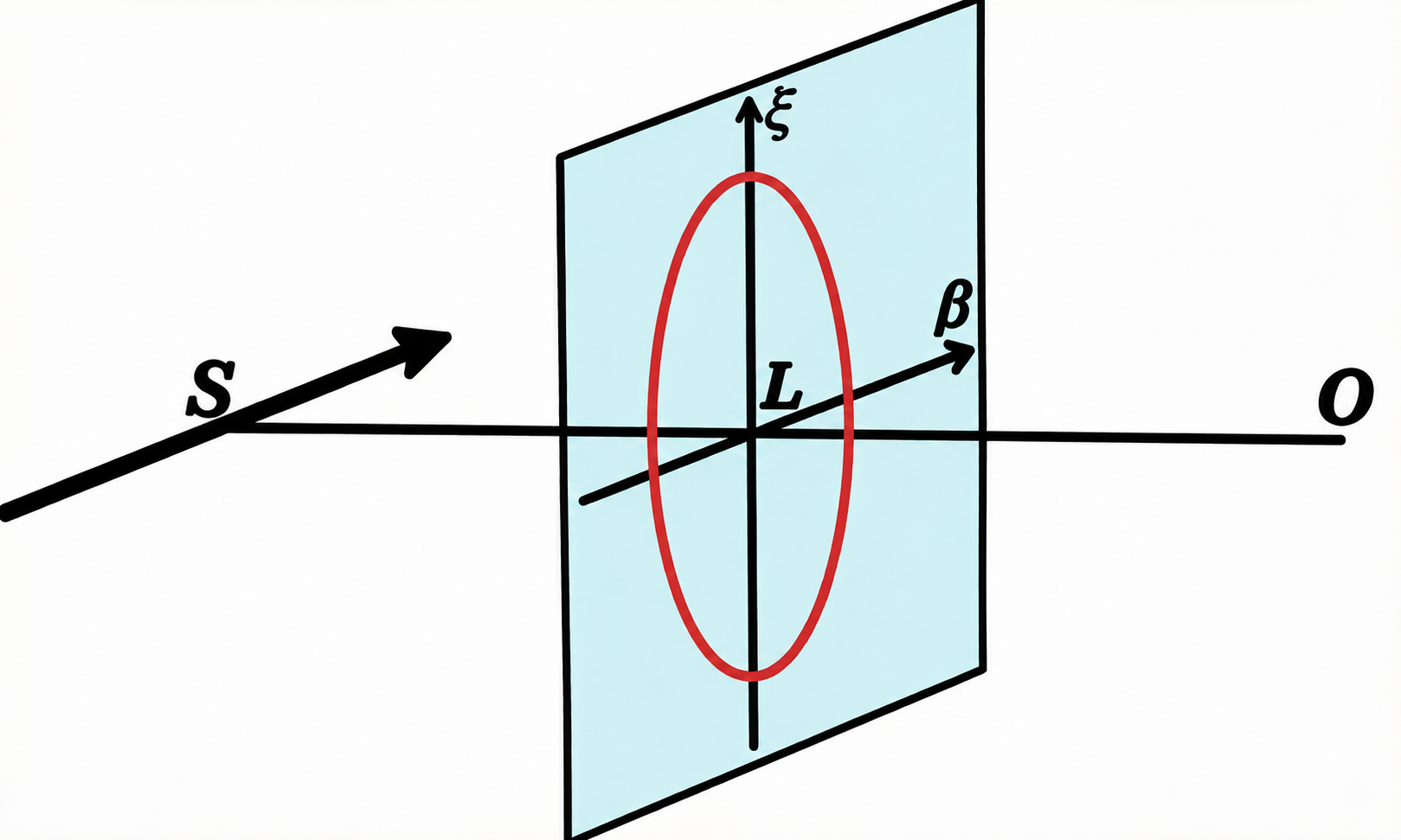}
	\caption{Sign convention in the lens plane. The observer-lens line defines the optical axis, and the lens is placed at the origin of the lens plane. The signed lens-plane coordinate is $\xi = D_{L} \theta$. Positive and negative values of $\xi$, or equivalently of $\theta$, indicate images formed on opposite sides of the optical axis. The closest-approach coordinate $r$ and the impact parameter $b_{\rm imp}$ are positive distances and depend on $|\theta|$, not on the sign of $\theta$. Therefore, an image with $\theta<0$ is not located at a negative radial distance; it is simply the image on the opposite side of the lens.}
	\label{FIG:2}
\end{figure}

Note that equation (27) is valid only for the positive coordinate. The signed lens equation is therefore
\begin{equation}
\beta = \frac{\xi}{|\xi|} \left[ \frac{b_{\rm imp} (|\xi|)}{D_{L}} - \frac{D_{LS}}{D_{S}} \widehat{a}(|\xi|)\right].
\end{equation}%

Substitution into Eq. (28), together with $r = |\xi|$, gives two different equations
\begin{eqnarray}
\beta &=& \frac{\xi}{D_{L}} \left[ 1 + \frac{\left(1-\gamma \right)r_{0}}{2\xi} 
+ \frac{\left(1 - 4\gamma + 3\gamma^{2}\right) r_{0}^{2}}{8\xi^{2}}\right] 
- \frac{D_{LS}}{D_{S}} \left[\frac{2 \left(1 - \gamma\right)r_{0}}{\xi} 
+ \frac{\pi \left( 11-22\gamma +15\gamma^{2}\right) r_{0}^{2}}{16\xi^{2}} \right. \nonumber \\
&& \left.- \frac{\left(1-\gamma \right)^{2} r_{0}^{2}}{\xi^{2}} \right], \quad {\rm for} \quad \xi > 0,
\end{eqnarray}
and
\begin{eqnarray}
\beta &=& \frac{\xi}{D_{L}} \left[ 1 - \frac{\left(1-\gamma \right)r_{0}}{2\xi} 
+ \frac{\left(1 - 4\gamma + 3\gamma^{2}\right) r_{0}^{2}}{8\xi^{2}}\right] 
- \frac{D_{LS}}{D_{S}} \left[\frac{2 \left(1 - \gamma\right)r_{0}}{\xi} 
- \frac{\pi \left( 11-22\gamma +15\gamma^{2}\right) r_{0}^{2}}{16\xi^{2}} \right. \nonumber \\
&& \left.+ \frac{\left(1-\gamma \right)^{2} r_{0}^{2}}{\xi^{2}} \right], \quad {\rm for} \quad \xi < 0.
\end{eqnarray}
These two equations are related by the symmetry $\beta (-\xi) = - \beta (\xi)$.

If the light source and the lens are exactly aligned with the line of sight (i.e. $\beta =0$), the image is expected to be round (Einstein ring). Using the condition $\beta =0$ in Eq.(29) we get
\begin{equation}
\xi^{3} + \frac{(1 - \gamma) r_{0}}{2} \xi^{2} 
-\frac{\left(1 - \gamma\right) \left\{16 D_{L} D_{LS} - \left(1 - 3\gamma \right) D_{S} r_{0} \right\}r_{0}}{8 D_{S}} \xi
- \frac{\left\{ \pi \left(11 - 22\gamma + 15\gamma
^{2} \right) - 16 \left(1 - \gamma\right)^{2} \right\} D_{L} D_{LS} r_{0}^{2}}{16D_{S}} = 0.
\end{equation}
For perfect alignment, the positive Einstein-ring radius is $R_{E} = |\xi|$. It is obtained by solving the cubic equation (29) and taking the positive root
\begin{equation}
R_{E} = \frac{1}{3}\sqrt{\frac{r_{0}\left( 1-\gamma \right) \left\{
48D_{L}D_{LS} - \left( 1-7\gamma \right) D_{S}r_{0} \right\} }{2D_{S}}}\sin %
\left( \frac{\pi }{2} - \frac{1}{3}\arccos {\Phi _{1}}\right) - \frac{\left( 1-\gamma \right) r_{0}}{6},
\end{equation}%
where 
\begin{equation*}
\Phi _{1} = \sqrt{\frac{D_{S}r_{0}}{2}} 
\frac{(1 - \gamma)^{2}(5 - 23\gamma) D_{S}r_{0} + 9\left\{3\pi \left(11 - 22\gamma + 15\gamma^{2} \right) - 64(1 - \gamma)^{2}\right\} D_{L}D_{LS} }{\left[\left(1 - \gamma \right) \left\{ 48D_{L}D_{LS}- \left(1 - 7\gamma\right) D_{S}r_{0} \right\} \right] ^{3/2}}.
\end{equation*}
The Einstein radius for a massless LPR wormhole can be obtained by solving the Eq.(31) after substituting $\gamma=1$
\begin{equation}
R_E = \left( \frac{\pi r_0^2}{4} \frac{D_LD_{LS}}{D_S} \right)^{1/3}.
\end{equation}

\begin{table}
\caption{The Einstein radius $R_{E}$ and its angular radius $\theta _{E}$ for the lens configuration when the source star
(source stars) is in the Galactic Bulge or in the LMC, and the LPR wormhole lens $(L)$ is at some intermediate distance (it is assumed that $D_{L}=4$ kpc and $D_{S}=8$ kpc for Bulge and $D_{L}=25$ kpc and $D_{S}=50$ kpc for LMC), so that $R_{E}$ is real.}
\begin{tabular}{|c|c|c|c||c|c|}
\hline
$r_{0}$ [km] & $\gamma$ & \multicolumn{2}{|c||}{Bulge} & \multicolumn{2}{|c|}{LMC
} \\ \cline{3-4}\cline{5-6}
&  & $R_{E}$ [km] & $\theta _{E}$ [mas] & $R_{E}$ [km] & $\theta _{E}$ [mas]
\\ \hline
$10^{-5}$ & $-0.95$ & 1.55123$\times 10^{6}$ & 0.0025929 & 3.87806$\times
10^{6}$ & 0.00103716 \\ 
$10^{-5}$ & $-0.5$ & 1.36051$\times 10^{6}$ & 0.00227412 & 3.40129$\times
10^{6}$ & 0.000909648 \\ 
$10^{-5}$ & $-0.05$ & 1.13829$\times 10^{6}$ & 0.00190267 & 2.84572$\times
10^{6}$ & 0.000761066 \\ 
$10^{-5}$ & $1$ & $1.69131\times 10^{2}$ & $0.0000003$ & $3.11542\times
10^{2}$ & $<0.0000001$ \\ \hline
$10^{-4}$ & $-0.95$ & 4.90541$\times 10^{6}$ & 0.00819946 & 1.22635$\times
10^{7}$ & 0.00327978 \\ 
$10^{-4}$ & $-0.5$ & 4.30232$\times 10^{6}$ & 0.0071914 & 1.07558$\times
10^{7}$ & 0.00287656 \\ 
$10^{-4}$ & $-0.05$ & 3.59958$\times 10^{6}$ & 0.00601676 & 8.99896$\times
10^{6}$ & 0.0024067 \\ 
$10^{-4}$ & $1$ & $7.85037\times 10^{2}$ & $0.000001$ & $1.44605\times 10^{3}
$ & $0.0000004$ \\ \hline
$10^{-3}$ & $-0.95$ & 1.55123$\times 10^{7}$ & 0.025929 & 3.87806$\times
10^{7}$ & 0.0103716 \\ 
$10^{-3}$ & $-0.5$ & 1.36051$\times 10^{7}$ & 0.0227412 & 3.40129$\times
10^{7}$ & 0.00909648 \\ 
$10^{-3}$ & $-0.05$ & 1.13829$\times 10^{7}$ & 0.0190267 & 2.84572$\times
10^{7}$ & 0.00761066 \\ 
$10^{-3}$ & $1$ & $3.64382\times 10^{3}$ & $0.000006$ & $6.71197\times 10^{3}
$ & $0.000002$ \\ \hline
$10^{-2}$ & $-0.95$ & 4.90541$\times 10^{7}$ & 0.0819946 & 1.22635$\times
10^{8}$ & 0.0327978 \\ 
$10^{-2}$ & $-0.5$ & 4.30232$\times 10^{7}$ & 0.071914 & 1.07558$\times
10^{8}$ & 0.0287656 \\ 
$10^{-2}$ & $-0.05$ & 3.59958$\times 10^{7}$ & 0.0601676 & 8.99896$\times
10^{7}$ & 0.024067 \\ 
$10^{-2}$ & $1$ & $1.69131\times 10^{4}$ & $0.00003$ & $3.11542\times 10^{4}$
& $0.000008$ \\ \hline
$10^{-1}$ & $-0.95$ & 1.55123$\times 10^{8}$ & 0.25929 & 3.87806$\times
10^{8}$ & 0.103716 \\ 
$10^{-1}$ & $-0.5$ & 1.36051$\times 10^{8}$ & 0.227412 & 3.40129$\times
10^{8}$ & 0.0909648 \\ 
$10^{-1}$ & $-0.05$ & 1.13829$\times 10^{8}$ & 0.190267 & 2.84572$\times
10^{8}$ & 0.0761066 \\ 
$10^{-1}$ & $1$ & $7.85037\times 10^{4}$ & $0.00013$ & $1.44605\times 10^{5}$
& $0.00004$ \\ \hline
$10^{0}$ & $-0.95$ & 4.90541$\times 10^{8}$ & 0.819946 & 1.22635$\times
10^{9}$ & 0.327978 \\ 
$10^{0}$ & $-0.5$ & 4.30232$\times 10^{8}$ & 0.71914 & 1.07558$\times 10^{9}$
& 0.287656 \\ 
$10^{0}$ & $-0.05$ & 3.59958$\times 10^{8}$ & 0.601676 & 8.99896$\times
10^{8}$ & 0.24067 \\ 
$10^{0}$ & $1$ & $3.64382\times 10^{5}$ & $0.00061$ & $6.71197\times 10^{5}$
& $0.00018$ \\ \hline
$10^{1}$ & $-0.95$ & 1.55123$\times 10^{9}$ & 2.5929 & 3.87806$\times 10^{9}$
& 1.03716 \\ 
$10^{1}$ & $-0.5$ & 1.36051$\times 10^{9}$ & 2.27412 & 3.40129$\times 10^{9}$
& 0.909648 \\ 
$10^{1}$ & $-0.05$ & 1.13829$\times 10^{9}$ & 1.90267 & 2.84572$\times 10^{9}
$ & 0.761066 \\ 
$10^{1}$ & $1$ & $1.69131\times 10^{6}$ & $0.00283$ & $3.11542\times 10^{6}$
& $0.00083$ \\ \hline
$10^{2}$ & $-0.95$ & 4.90541$\times 10^{9}$ & 8.19946 & 1.22635$\times
10^{10}$ & 3.27978 \\ 
$10^{2}$ & $-0.5$ & 4.30232$\times 10^{9}$ & 7.1914 & 1.07558$\times 10^{10}$
& 2.87656 \\ 
$10^{2}$ & $-0.05$ & 3.59958$\times 10^{9}$ & 6.01676 & 8.99896$\times 10^{9}
$ & 2.4067 \\ 
$10^{2}$ & $1$ & $7.85037\times 10^{6}$ & $0.01314$ & $1.44605\times 10^{7}$
& $0.00387$ \\ \hline
$10^{3}$ & $-0.95$ & 1.55123$\times 10^{10}$ & 25.929 & 3.87806$\times
10^{10}$ & 10.3716 \\ 
$10^{3}$ & $-0.5$ & 1.36051$\times 10^{10}$ & 22.7412 & 3.40129$\times
10^{10}$ & 9.09648 \\ 
$10^{3}$ & $-0.05$ & 1.13829$\times 10^{10}$ & 19.0267 & 2.84572$\times
10^{10}$ & 7.61067 \\ 
$10^{3}$ & $1$ & $3.64382\times 10^{7}$ & $0.06101$ & $6.71197\times 10^{7}$
& $0.01798$ \\ \hline
$10^{4}$ & $-0.95$ & 4.90541$\times 10^{10}$ & 81.9946 & 1.22635$\times
10^{11}$ & 32.7978 \\ 
$10^{4}$ & $-0.5$ & 4.30233$\times 10^{10}$ & 71.914 & 1.07558$\times 10^{11}
$ & 28.7656 \\ 
$10^{4}$ & $-0.05$ & 3.59958$\times 10^{10}$ & 60.1676 & 8.99896$\times
10^{10}$ & 24.067 \\ 
$10^{4}$ & $1$ & $1.69131\times 10^{8}$ & $0.28316$ & $3.11542\times 10^{8}$
& $0.08345$ \\ \hline
$10^{5}$ & $-0.95$ & 1.55123$\times 10^{11}$ & 259.29 & 3.87806$\times
10^{11}$ & 103.716 \\ 
$10^{5}$ & $-0.5$ & 1.36052$\times 10^{11}$ & 227.412 & 3.40129$\times
10^{11}$ & 90.9649 \\ 
$10^{5}$ & $-0.05$ & 1.13829$\times 10^{11}$ & 190.267 & 2.84572$\times
10^{11}$ & 76.1067 \\ 
$10^{5}$ & $1$ & $7.85037\times 10^{8}$ & $1.31433$ & $1.44605\times 10^{9}$
& $0.38736$ \\ \hline
\end{tabular}
\end{table}

Using the substitution of positive image with $\theta_{1} = \xi/D_{L} > 0$ and negative image with $\theta_{2} = \xi/D_{L} < 0$, Eqs.(29) and (30) can be rewritten as
\begin{eqnarray}
\beta &=& \theta_{1} + \frac{\left(1 - \gamma \right) r_{0}}{2D_{L}} + \frac{\left\{\left(1 - 4\gamma + 3\gamma^{2} \right) D_{S}r_{0} - 16 \left(1-\gamma \right) D_{LS} D_{L} \right\} r_{0}}{8 D_{S} D_{L}^{2}} \frac{1}{\theta_{1}} + \frac{\left(1 - \gamma\right)^{2} D_{LS} r_{0}^{2}}{D_{S}D_{L}^{2}} \frac{1}{\theta_{1}^{2}} \nonumber \\
&&- \frac{\pi \left( 11-22\gamma+15\gamma^{2}\right) D_{LS} r_{0}^{2}}{16 D_{S}D_{L}^{2}} \frac{1}{\theta_{1}^{2}}, \quad {\rm for} \quad \theta_{1} > 0,
\end{eqnarray}
\begin{eqnarray}
\beta &=& \theta_{2} - \frac{\left(1 - \gamma \right) r_{0}}{2D_{L}} + \frac{\left\{\left(1 - 4\gamma + 3\gamma^{2} \right) D_{S}r_{0} - 16 \left(1-\gamma \right) D_{LS} D_{L} \right\} r_{0}}{8 D_{S} D_{L}^{2}} \frac{1}{\theta_{2}} - \frac{\left(1 - \gamma\right)^{2} D_{LS} r_{0}^{2}}{D_{S}D_{L}^{2}} \frac{1}{\theta_{2}^{2}} \nonumber \\
&&+ \frac{\pi \left( 11-22\gamma+15\gamma^{2}\right) D_{LS} r_{0}^{2}}{16 D_{S}D_{L}^{2}} \frac{1}{\theta_{2}^{2}},  \quad {\rm for} \quad \theta_{2} < 0.
\end{eqnarray}

Using the angular Einstein radius $\theta_{E} =R_{E}/D_{L}$, we obtain
\begin{eqnarray}
\beta &=& \theta_{1} + \frac{\left(1 - \gamma \right) r_{0}\theta_{E}}{2R_{E}} + \frac{\left\{\left(1 - 4\gamma + 3\gamma^{2} \right) D_{S}r_{0} - 16 \left(1-\gamma \right) D_{LS} D_{L} \right\} r_{0}}{8 D_{S}R_{E}^{2}} \frac{\theta_{E}^2}{\theta_{1}} + \frac{\left(1 - \gamma\right)^{2} D_{LS} r_{0}^{2}}{D_{S}R_{E}^{2}} \frac{\theta_{E}^2}{\theta_{1}^{2}} \nonumber \\
&& - \frac{\pi \left( 11-22\gamma+15\gamma^{2}\right) D_{LS} r_{0}^{2}}{16 D_{S}R_{E}^{2}} \frac{\theta_{E}^2}{\theta_{1}^{2}}, \quad {\rm for} \quad \theta_{1} > 0,
\end{eqnarray}

\begin{eqnarray}
\beta &=& \theta_{2} - \frac{\left(1 - \gamma \right) r_{0}\theta_{E}}{2R_{E}} + \frac{\left\{\left(1 - 4\gamma + 3\gamma^{2} \right) D_{S}r_{0} - 16 \left(1-\gamma \right) D_{LS} D_{L} \right\} r_{0}}{8 D_{S}R_{E}^{2}} \frac{\theta_{E}^2}{\theta_{2}} - \frac{\left(1 - \gamma\right)^{2} D_{LS} r_{0}^{2}}{D_{S}R_{E}^{2}} \frac{\theta_{E}^2}{\theta_{2}^{2}} \nonumber \\
&& + \frac{\pi \left( 11-22\gamma+15\gamma^{2}\right) D_{LS} r_{0}^{2}}{16 D_{S}R_{E}^{2}} \frac{\theta_{E}^2}{\theta_{2}^{2}}, \quad {\rm for} \quad \theta_{2} < 0.
\end{eqnarray}
Eqs. (36), (37) can be rewritten as
\begin{eqnarray}
&&\frac{\theta_{1}^{3}}{\theta _{E}^{3}}-\frac{\beta }{\theta _{E}}\frac{\theta_{1}^{2}}{\theta _{E}^{2}} + \frac{\left(1-\gamma \right) r_{0}}{2R_{E}} \frac{\theta_{1}^{2}}{\theta_{E}^{2}} + \frac{\left\{\left(1 - 4\gamma + 3\gamma^{2} \right)D_{S}r_{0} - 16 \left(1-\gamma \right) D_{LS} D_{L} \right\} r_{0}}{8 D_{S} R_{E}^{2}} \frac{\theta_{1}}{\theta_{E}} \nonumber \\
&& + \frac{\left(1 - \gamma\right)^{2} D_{L} D_{LS} r_{0}^{2}}{D_{S}R_{E}^{3}} - \frac{\pi \left( 11-22\gamma+15\gamma^{2}\right) r_{0}^{2}}{16}\frac{D_{L}D_{LS}}{%
D_{S}R_{E}^{3}} = 0, \quad {\rm for} \quad \theta_{1} > 0,
\end{eqnarray}

\begin{eqnarray}
&&\frac{\theta_{2}^{3}}{\theta _{E}^{3}}-\frac{\beta }{\theta _{E}}\frac{\theta_{2}^{2}}{\theta _{E}^{2}} - \frac{\left(1-\gamma \right) r_{0}}{2R_{E}} \frac{\theta_{2}^{2}}{\theta_{E}^{2}} + \frac{\left\{\left(1 - 4\gamma + 3\gamma^{2} \right)D_{S}r_{0} - 16 \left(1-\gamma \right) D_{LS} D_{L} \right\} r_{0}}{8 D_{S} R_{E}^{2}} \frac{\theta_{2}}{\theta_{E}} \nonumber \\
&& - \frac{\left(1 - \gamma\right)^{2} D_{L} D_{LS} r_{0}^{2}}{D_{S}R_{E}^{3}} + \frac{\pi \left( 11-22\gamma+15\gamma^{2}\right) r_{0}^{2}}{16}\frac{D_{L}D_{LS}}{%
D_{S}R_{E}^{3}} = 0, \quad {\rm for} \quad \theta_{2} < 0.
\end{eqnarray}

Using reduced parameters $\hat{\beta} = \beta /\theta_{E}$ and $\hat{\theta} = \theta /\theta_{E}$, Eqs. (38), (39) becomes a cubic equations: 
\begin{equation}
\hat{\theta_{1}}^{3}-\left(\hat{\beta} - P\right)\hat{\theta_{1}}^{2}+M\hat{\theta_{1}}+N=0, \quad {\rm for} \quad \hat{\theta_{1}} > 0,
\end{equation}
\begin{equation}
\hat{\theta_{2}}^{3}-\left(\hat{\beta} + P\right)\hat{\theta_{2}}^{2}+M\hat{\theta_{2}}-N=0, \quad {\rm for} \quad \hat{\theta_{2}} < 0,
\end{equation}%
where $P$, $M$ and $N$ are functions of $\gamma$, $r_{0}$, $D_{S}$, $D_{L}$, $D_{LS}$ and given as 
\begin{eqnarray}
P &=& \frac{\left(1-\gamma \right) r_{0}}{2R_{E}}, \nonumber \\
M &=& \frac{\left\{\left(1 - 4\gamma + 3\gamma^{2} \right)D_{S}r_{0} - 16 \left(1-\gamma \right) D_{LS} D_{L} \right\} r_{0}}{8 D_{S} R_{E}^{2}}, \nonumber \\
N &=& \frac{\left(1 - \gamma\right)^{2} D_{L} D_{LS} r_{0}^{2}}{D_{S}R_{E}^{3}} - \frac{\pi \left( 11-22\gamma+15\gamma^{2}\right) r_{0}^{2}}{16}\frac{D_{L}D_{LS}}{%
D_{S}R_{E}^{3}}. \nonumber
\end{eqnarray}%

Solving the cubic equation (40), we obtain one positive root, which denotes the positive image
\begin{equation}
\widehat{\theta}_{1} = \frac{P-\hat{\beta}}{3} + \frac{2\sqrt{\left(P-\hat{\beta}\right)^{2} - 3M}}{3} \sin{\left[\frac{\pi}{6} + \frac{1}{3} \arccos{\left\{\frac{2\left(P-\hat{\beta}\right)^{3} - 9M\left(P-\hat{\beta}\right) - 27N}{2\left(\left(P-\hat{\beta}\right)^{2}-3M\right)^{3/2}} \right\}} \right]}, \quad \left(
\hat{\theta}_{1} > 0\right),
\end{equation}

Solving the cubic equation (41), we obtain one negative root, which denotes the negative image
\begin{equation}
\widehat{\theta}_{2} = \frac{P+\hat{\beta}}{3} - \frac{2\sqrt{\left(P+\hat{\beta}\right)^{2} - 3M}}{3} \cos{\left[\frac{1}{3} \arccos{\left\{-\frac{2\left(P+\hat{\beta}\right)^{3} - 9M\left(P-\hat{\beta}\right) + 27N}{2\left(\left(P+\hat{\beta}\right)^{2}-3M\right)^{3/2}} \right\}} \right]}, \quad \left(\hat{\theta}_{2} < 1\right).
\end{equation}

Fig. 3 illustrates the dependence of the image positions, $\widehat{\theta}_{1}$ and $\widehat{\theta}_{2}$, on the source position $\hat{\beta}$. It is evident that, for $\hat{\beta}\geq 0$, the primary image $\widehat{\theta}_{1}$ is located outside the Einstein ring, whereas the secondary image $\widehat{\theta}_{2}$ lies inside the Einstein ring. Conversely, when $\hat{\beta}< 0$, the image configuration is reversed: $\widehat{\theta}_{2}$ is positioned outside the Einstein ring, while $\widehat{\theta}_{1}$ is located inside it.

\begin{figure}
	\centering
	\includegraphics[width=.75\textwidth]{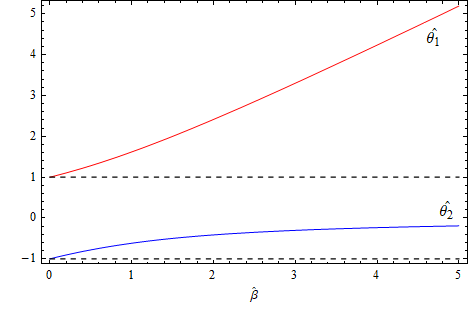}
	\caption{Reduced image positions for the LPR wormhole as functions of the reduced source position $\hat{\beta}$. Both axes are dimensionless. The dashed lines show the reduced Einstein-ring radii $\hat{\theta}=\pm1$. The branch $\hat{\theta}_1$ corresponds to the outer image, while $\hat{\theta}_2$ corresponds to the inner image.}
	\label{FIG:2}
\end{figure}

We use the Paczy\'{n}ski expression \cite{Paczynski:1986} for magnification of a brightness $A$ of the light curves: 
\begin{equation}
A=A_{1}+A_{2}=\left\vert \frac{\widehat{\theta }_{1}}{\widehat{\beta }}\frac{%
d\widehat{\theta }_{1}}{d\widehat{\beta }}\right\vert +\left\vert \frac{%
\widehat{\theta }_{2}}{\widehat{\beta }}\frac{d\widehat{\theta }_{2}}{d%
\widehat{\beta }}\right\vert ,
\end{equation}%
where $A_{1}$, $A_{2}$ are the magnification of the external and internal images. Since the analytical expressions obtained in Eq.(44) are very large, we will present the results graphically.

The lens movement can be described by time dependence as follows \cite{Abe:2010} 
\begin{equation}
\widehat{\beta }=\sqrt{\widehat{\beta }_{0}^{2}+\frac{\left( t-t_{0}\right)
^{2}}{t_{E}^{2}}},
\end{equation}%
where $\widehat{\beta }_{0}$ is the impact parameter of the source trajectory, $t_{0}$ is the time of closest approach, $t_{E}$ is the Einstein radius crossing time, given by
\begin{equation}
t_{E}=\frac{R_{E}}{\vartheta _{T}},
\end{equation}%
where $\vartheta _{T}$ is the transverse velocity of the lens relative to
the source and the observer.

\begin{figure}
	\centering
\includegraphics[width=1.0\textwidth]{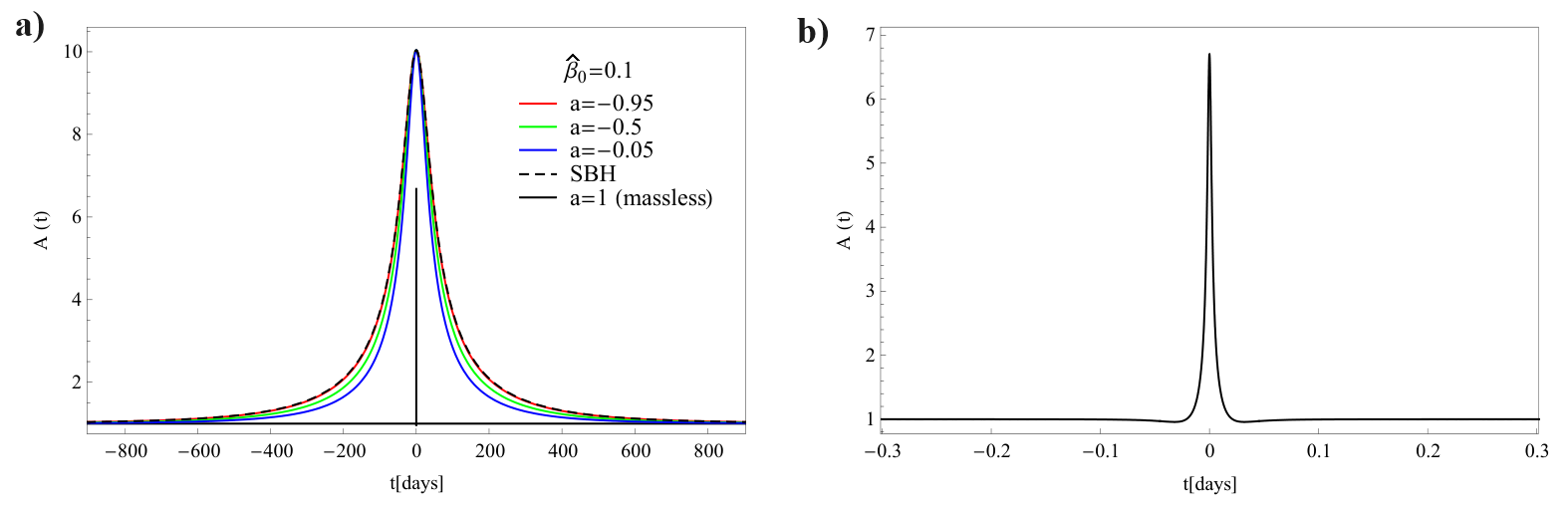}
	\caption{Idealized point-source Paczy\'{n}ski light curves for the massive LPR wormhole branch $-1<\gamma<0$ ($\gamma=-0.95$ -- solid red line, $\gamma=-0.5$ -- solid green line, $\gamma=-0.05$ -- solid blue line) and for the massless comparison case $\gamma=1$ (solid black line), with $\hat{\beta}_0 = 0.1$. The vertical axis is the dimensionless total magnification $A(t)$, and the horizontal axis is time in seconds, with $t_0 = 0$. The parameters are $r_0=1\,{\rm km}$, $D_L=4\,{\rm kpc}$, $D_S=8\,{\rm kpc}$, and $v_T=220\,{\rm km\,s^{-1}}$, unless otherwise stated. The Schwarzschild curve, when shown, is a separate reference lens with the same ADM mass and is not obtained by setting $\gamma=0$ in the LPR metric.}
	\label{FIG:3}
\end{figure}

\begin{figure}
	\centering
\includegraphics[width=1.0\textwidth]{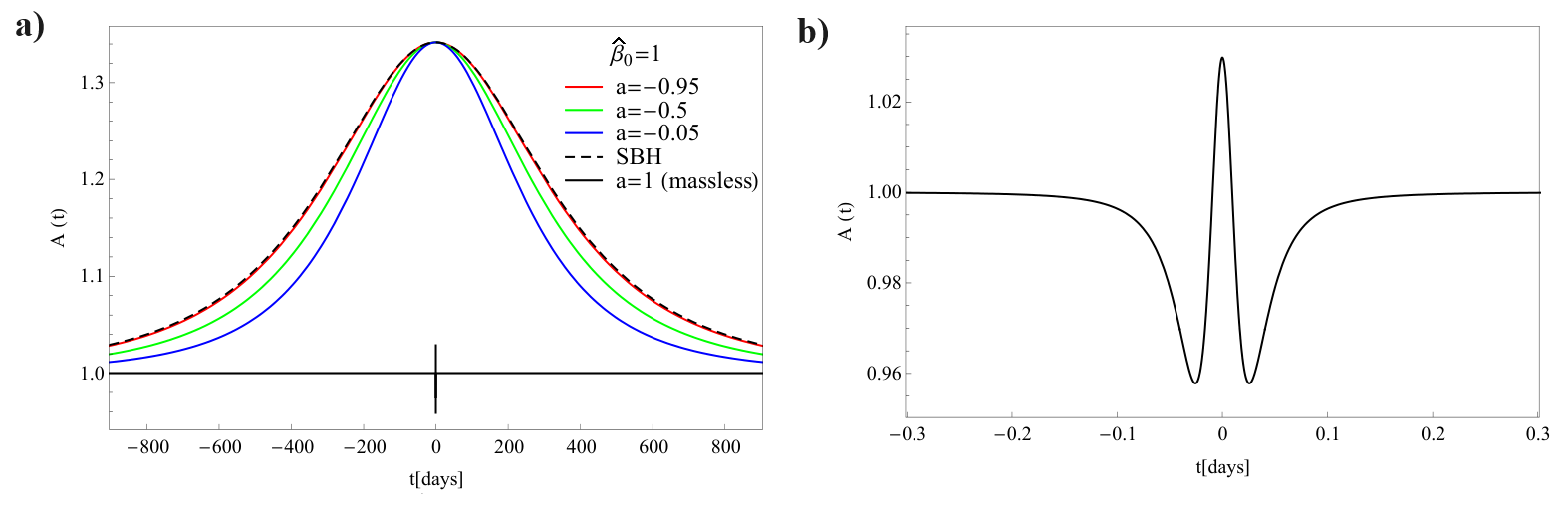}
	\caption{Idealized point-source Paczy\'{n}ski light curves for the massive LPR wormhole branch $-1<\gamma<0$ ($\gamma=-0.95$ -- solid red line, $\gamma=-0.5$ --
solid green line, $\gamma=-0.05$ -- solid blue line) and for the massless comparison case $\gamma=1$ (solid black line), with $\hat{\beta}_0 = 1$. The axes and units are the same as in Fig. 4.}
	\label{FIG:4}
\end{figure}

The light curves in Figs. 4 and 5 are idealized point-source light curves. They are useful for displaying the qualitative difference between the massive LPR branch and the massless comparison case, but they should not be interpreted as realistic survey light curves unless $\rho_{\star} = \theta_\star / \theta_E  \ll 1$, where $\theta_\star$ is the physical angular radius of the source star.

Figs.4 a-b show the light curves of LPR phantom wormhole for different
values of the parameter $\hat{\beta}_{0}=0.1$ (a -- massive LPR case), (b --
massless LPR case) for various values of $\gamma$, denoted as
follows: $\gamma=-0.95$ (solid red line), $\gamma=-0.5$ (solid green line), $\gamma=-0.05$
(solid blue line) and $\gamma=1$ corresponds massless LPR wormhole (solid black
line). It is evident from Fig.4a that the peak of the image brightness,
which occurs at $t=0$, is the same for any values of the parameter $\gamma$ of
the massive wormhole of the LPR. The difference appears at a small deviation
from the peak point on both sides, that is, at values of $|t|$ from $0$ to $%
800$ days. The greatest difference is achieved at $|t|=50$ days. Note that
the light curves for the massless LPR wormhole greatly differ from the massive
case. The main difference appears near the peak, where the peak of the
massive wormhole brightness is $1.5$ times greater than the peak of the
massless wormhole. Moreover, for the massless case, the light curves exhibit
gutters\footnote{In what follows, "gutters" denotes a demagnification trough, when $A(t)<1$, in the total point-source magnification. Such troughs are characteristic of massless wormhole lensing, where the first nonzero deflection term is proportional to $1/r^2$. In contrast, for the massive LPR branch $-1<\gamma<0$, the leading deflection is proportional to $1/r$, and the point-source light curves are Paczy\'{n}ski-like.}. However, all this occurs in a very short time interval $|t|$ from $%
0 $ to $0.1$ days, which is $8000$ times less than the massive case. A
similar picture is observed in Fig.4b. Note that in this case the peak
brightness of the massive wormhole is $1.3$ times larger than the peak of
the massless wormhole. The magnitude of the brightness change of the
massless wormhole varies in the time interval $|t|$ from $0$ to $0.2$ days,
which is $8000$ times smaller compared to the massive case.

Figs.5 a-b show the light curves of LPR phantom wormhole for different values of the parameter $\hat{\beta}_{0} = 1$ (a -- massive LPR case), (b -- massless LPR case) for various values of the $\gamma$ parameter, denoted as follows: $\gamma = -0.95$ (solid red line), $\gamma=-0.5$ (solid green line), $\gamma = -0.05$ (solid blue line) and $\gamma = 1$ corresponds massless LPR wormhole (solid black line). In Figure 5, we see similar behavior of the light curves for the LPR wormhole. With increasing $\hat{\beta}_{0}$, a decrease in peaks is observed for both the massive and massless wormholes. Increasing $\hat{\beta}_{0}$ leads to an increase in gutters, typical of massless wormholes.

The point-source approximation is valid only when
\begin{equation}
\rho_\star=\frac{\theta_\star}{\theta_E} = \frac{D_L R_\star}{D_S R_E} \ll 1, 
\end{equation}
where $R_\star$ is the physical radius of the source star. If $\rho_\star\gtrsim1$, finite-source averaging suppresses sharp peaks and gutters, and the point-source light curves shown in Figs. 4 and 5 are no longer reliable. Therefore, detectability statements for small $R_E$ should be read with this limitation in mind.

Table 2 presents numerical estimates of $t_{E}$ for the Bulge and LMC. With
numerical values of $r_{0}$ from $1$ km to $10^{9}$ km and values of $\gamma$
from $-0.95$ to $-0.05$.

\begin{table}
\caption{The Einstein radius crossing time $t_{E}$ for the lens configuration when the source star (source stars) is in the Galactic Bulge and in LMC, and the LPR wormhole lens $(L)$ is at some intermediate distance.}
\begin{tabular}{|c|c|c|c||c|c|c|c|}
\hline
$r_{0}$ [km] & $\gamma $ & \multicolumn{2}{|c||}{$t_{E}$ [days]} & $r_{0}$
[km] & $\gamma $ & \multicolumn{2}{|c|}{$t_{E}$ [days]} \\ 
\cline{3-4}\cline{7-8}
&  & Bulge & LMC &  &  & Bulge & LMC \\ \hline
$10^{-5}$ & $-0.95$ & $0.08161$ & $0.20402$ &
$10^{1}$ & $-0.95$ & $81.60908$ & $204.02269$ \\ 
$10^{-5}$ & $-0.5$  & $0.07158$ & $0.17894$ &
$10^{1}$ & $-0.5$  & $71.57589$ & $178.93974$ \\ 
$10^{-5}$ & $-0.05$ & $0.05988$ & $0.14971$ &
$10^{1}$ & $-0.05$ & $59.88469$ & $149.71172$ \\ 
$10^{-5}$ & $1$     & $0.00001$ & $0.00002$ &
$10^{1}$ & $1$     & $0.08898$ & $0.16390$ \\
\hline
$10^{-4}$ & $-0.95$ & $0.25807$ & $0.64518$ & $10^{2}$ & $-0.95$ & $258.07056$ & $645.17640$ \\ 
$10^{-4}$ & $-0.5$ & $0.22634$ & $0.56586$ & $10^{2}$ & $-0.5$ & $226.34285$ & $565.85713$ \\ 
$10^{-4}$ & $-0.05$ & $0.18937$ & $0.47343$ & $10^{2}$ & $-0.05$ & $189.37202$ & $473.43004$ \\ 
$10^{-4}$ & $1$ & $0.00004$ & $0.00008$ & $10^{2}$ & $1$ & $0.41300$ & $0.76076$ \\
\hline
$10^{-3}$ & $-0.95$ & $0.81609$ & $2.04023$ & $10^{3}$ & $-0.95$ & $816.09079$ & $2040.22694$ \\ 
$10^{-3}$ & $-0.5$ & $0.71576$ & $1.78940$ & $10^{3}$ & $-0.5$ & $715.75897$ & $1789.39739$ \\ 
$10^{-3}$ & $-0.05$ & $0.59885$ & $1.49712$ & $10^{3}$ & $-0.05$ & $598.84691$ & $1497.11727$ \\ 
$10^{-3}$ & $1$ & $0.00019$ & $0.00035$ & $10^{3}$ & $1$ & $1.91699$ & $3.53113$ \\
\hline
$10^{-2}$ & $-0.95$ & $2.58071$ & $6.45176$ & $10^{4}$ & $-0.95$ & $2580.70593$ & $6451.76432$ \\ 
$10^{-2}$ & $-0.5$ & $2.26343$ & $5.65857$ & $10^{4}$ & $-0.5$ & $2263.42878$ & $5658.57157$ \\ 
$10^{-2}$ & $-0.05$ & $1.89372$ & $4.73430$ & $10^{4}$ & $-0.05$ & $1893.72034$ & $4734.30060$ \\ 
$10^{-2}$ & $1$ & $0.00089$ & $0.00164$ & $10^{4}$ & $1$ & $8.89789$ & $16.39006$ \\
\hline
$10^{-1}$ & $-0.95$ & $8.16091$ & $20.40227$ & $10^{5}$ & $-0.95$ & $8160.91108$ & $20402.27252$ \\ 
$10^{-1}$ & $-0.5$ & $7.15759$ & $17.89397$ & $10^{5}$ & $-0.5$ & $7157.59194$ & $17893.97616$ \\ 
$10^{-1}$ & $-0.05$ & $5.98847$ & $14.97117$ & $10^{5}$ & $-0.05$ & $5988.47061$ & $14971.17412$ \\ 
$10^{-1}$ & $1$ & $0.00413$ & $0.00761$ & $10^{5}$ & $1$ & $41.30035$ & $76.07590$ \\
\hline
$10^{0}$ & $-0.95$ & $25.80706$ & $64.51764$ & $10^{6}$ & $-0.95$ & $25807.09041$ & $64517.67423$\\ 
$10^{0}$ & $-0.5$ & $22.63428$ & $56.58571$ & $10^{6}$ & $-0.5$ & $22634.30992$ & $56585.73789$ \\ 
$10^{0}$ & $-0.05$ & $18.93720$ & $47.34300$ & $10^{6}$ & $-0.05$ & $18937.21779$ & $47343.02042$ \\ 
$10^{0}$ & $1$ & $0.01917$ & $0.03531$ & $10^{6}$ & $1$ & $191.69926$ & $353.11306$ \\ \hline
\end{tabular}
\end{table}

%%%%%%%%%%%%%%%%%%%%%%%%%%%%%%%%%%%%%%%%%
\section{Probabilistic characteristics of microlensing}
\label{sec5}
%%%%%%%%%%%%%%%%%%%%%%%%%%%%%%%%%%%%%%%%%
The probability of a microlensing event to occur for a star is expressed by
the optical depth $\tau$ \cite{Abe:2010}: 
\begin{equation}
\tau =\pi \int\limits_{0}^{D_{S}}n(D_{L})R_{E}^{2}dD_{L},
\end{equation}%
where $n(D_{L})$ is the number density of wormholes as a function of the
line of sight. Here we simply assume that $n(D_{L})$ is constant ($n(D_{L})=n $) and $D_{LS} = D_{S} - D_{L}$: 
\begin{equation}
\tau =\frac{\pi n \left( 1-\gamma \right) r_{0}}{18} \int\limits_{0}^{D_{S}}{\left[\sqrt{\frac{
48D_{L}\left(D_{S} - D_{L}\right)-D_{S}r_{0}\left( 1-7\gamma \right)}{D_{S}}}\sin %
\left( \frac{\pi }{2} - \frac{1}{3}\arccos {\Phi _{1}}\right) - \sqrt{\frac{\left( 1-\gamma \right) r_{0}}{2}}\right]^{2}}dD_{L}.
\end{equation}

The event rate expected for a source star $\Gamma $ is defined by 
\begin{equation}
\Gamma = 2\int\limits_{0}^{D_{S}}n(D_{L})R_{E}v_{T} dD_{L}.
\end{equation}%
Using the Einstein radius, given by Eq.(32), with the massless comparison case given by Eq. (33), and assuming that $n(D_{L})=n$
is constant we can rewrite the above as
\begin{equation}
\Gamma = \frac{2nv_{T}}{3} \int\limits_{0}^{D_{S}} \left[\sqrt{\frac{r_{0}\left( 1-\gamma \right) \left\{ 48D_{L}\left(D_{S} - D_{L}\right) - D_{S}r_{0}\left( 1-7\gamma \right) \right\} }{2D_{S}}}\sin %
\left( \frac{\pi }{2} - \frac{1}{3}\arccos {\Phi _{1}}\right) -\frac{\left( 1-\gamma \right) r_{0}}{2} \right] dD_{L}.
\end{equation}

The line-of-sight integrals in Eqs. (49) and (51) cannot be evaluated analytically, so the corresponding numerical values of the optical depth $\tau$ and microlensing event rate $\Gamma$ of LPR wormhole are presented in Table 3.

As was pointed out by Abe \cite{Abe:2010} there is no reliable prediction of the number density of wormholes. Some authors \cite{Krasnikov:2000, Lobo:2008} have speculated that wormholes are very common in the universe, at least as abundant as stars. Even if we accept such speculation, there are still large uncertainties in the value of $n$ because the distribution of wormholes is not specified. In \cite{Abe:2010, Akhtaryanova:2024b}
authors considered two possibilities that wormholes are bound to the Galaxy and the number density is approximately equal to the local stellar density. The other possibility is that wormholes are not bound to the Galaxy and are approximately uniformly distributed throughout the universe. Therefore, the choice $n = \rho_{Ls}/\langle M_\star\rangle = 0.147\,{\rm pc}^{-3}$ should be regarded only as an illustrative normalization, not as a realistic Galactic population model. A realistic event-rate prediction would require a Galactic density profile, a velocity distribution, finite-source treatment, survey cadence, blending, and detection efficiency. The present calculation is intended to show the parametric dependence of $\tau$ and $\Gamma$ on $r_0$ and $\gamma$, rather than to provide robust detection thresholds.

Using these values, we calculated the optical depths and event rates for Bulge and LMC lensings presented in Table 3. In an ordinary Schwarzschild microlensing survey, observations are made of more than 10 million stars. Thus, we can expect approximately $10^{7}\Gamma $ events in a year. However, the situation is different in a wormhole search. As mentioned previously, the magnification of wormhole lensing is less than that of Schwarzschild lensing, and a remarkable feature of wormhole lensing is the decreasing brightness around the Einstein radius crossing times.

The values in Table 3 are illustrative estimates obtained under a simplified optically thin model with constant number density, fixed transverse velocity, and point-source light curves. They should not be interpreted as survey detection thresholds or as constraints on the abundance of LPR wormholes. Their purpose is only to show how $\tau$ and $\Gamma$ scale with the throat radius $r_{0}$ and the LPR parameter $\gamma$. A quantitative observational prediction would require a Galactic lens-density profile, a velocity distribution, finite-source light curves, blending, cadence, and detection efficiency.

If no candidate is found, we can set upper limits of $\Gamma $ and/or $\tau$ as functions of $t_{E}$. To convert these values to physical parameters ($n$ and $\gamma$) requires the distribution of $v_{T}$. At present, there is no reliable model of the distribution except for using the bound or unbound hypothesis. On the other hand, the event rates for the unbound model are too small for the events to be detected.

\begin{table}
\caption{Optical depth $\tau$ and event rate $\Gamma$ for source stars in the Galactic Bulge ($D_S=8\,{\rm kpc}$) and in the LMC ($D_S=50\,{\rm kpc}$). We use $v_T=220\,{\rm km\,s^{-1}}$ and the illustrative constant number density $n=0.147\,{\rm pc}^{-3}$. For numerical calculations the second order Einstein radius was used in the line-of-sight integrals. The values are illustrative and do not include finite-source effects, survey cadence, blending, or detection efficiency.}
\begin{tabular}{|c|c|c|c|c|c|}
\hline
$r_{0}(km)$ & $\gamma$ & \multicolumn{2}{|c|}{Bulge} & \multicolumn{2}{|c|}{LMC}
\\ \cline{3-6}
&  & $\tau $ & $\Gamma $ (1/year) & $\tau $ & $\Gamma $ (1/year) \\ \hline
$10^{-5}$ & $-0.95$ & $6.22738\ast 10^{-12}$ & 2.09464$\ast 10^{-8}$ & 
2.43257$\ast 10^{-10}$ & 3.27287$\ast 10^{-7}$ \\ 
$10^{-5}$ & $-0.5$ & $4.79029\ast 10^{-12}$ & 1.83712$\ast 10^{-8}$ & 1.87121%
$\ast 10^{-10}$ & 2.87050$\ast 10^{-7}$ \\ 
$10^{-5}$ & $-0.05$ & $3.3532\ast 10^{-12}$ & 1.53705$\ast 10^{-8}$ & 1.30985%
$\ast 10^{-10}$ & 2.40163$\ast 10^{-7}$ \\ 
$10^{-5}$ & $1$ & $8.21694\ast 10^{-20}$ & 2.4477$\ast 10^{-12}$ & 1.74252$%
\ast 10^{-18}$ & 2.81794$\ast 10^{-11}$ \\ \hline
$10^{-4}$ & $-0.95$ & $6.22738\ast 10^{-11}$ & 6.62383$\ast 10^{-8}$ & 
2.43257$\ast 10^{-9}$ & 1.03497$\ast 10^{-6}$ \\ 
$10^{-4}$ & $-0.5$ & $4.79029\ast 10^{-11}$ & 5.80948$\ast 10^{-8}$ & 1.87121%
$\ast 10^{-9}$ & 9.07732$\ast 10^{-7}$ \\ 
$10^{-4}$ & $-0.05$ & $3.3532\ast 10^{-11}$ & 4.86056$\ast 10^{-8}$ & 1.30985%
$\ast 10^{-9}$ & 7.59463$\ast 10^{-7}$ \\ 
$10^{-4}$ & $1$ & $1.77029\ast 10^{-18}$ & 1.13612$\ast 10^{-11}$ & 3.75414$%
\ast 10^{-17}$ & 1.30797$\ast 10^{-10}$ \\ \hline
$10^{-3}$ & $-0.95$ & $6.22738\ast 10^{-10}$ & 2.09464$\ast 10^{-7}$ & 
2.43257$\ast 10^{-8}$ & 3.27287$\ast 10^{-6}$ \\ 
$10^{-3}$ & $-0.5$ & $4.79029\ast 10^{-10}$ & 1.83712$\ast 10^{-7}$ & 1.87121%
$\ast 10^{-8}$ & 2.87050$\ast 10^{-6}$ \\ 
$10^{-3}$ & $-0.05$ & $3.3532\ast 10^{-10}$ & 1.53705$\ast 10^{-7}$ & 1.30985%
$\ast 10^{-8}$ & 2.40163$\ast 10^{-6}$ \\ 
$10^{-3}$ & $1$ & $3.81397\ast 10^{-17}$ & 5.27341$\ast 10^{-11}$ & 8.08805$%
\ast 10^{-16}$ & 6.07107$\ast 10^{-10}$ \\ \hline
$10^{-2}$ & $-0.95$ & $6.22738\ast 10^{-9}$ & 6.62383$\ast 10^{-7}$ & 2.43257%
$\ast 10^{-7}$ & 1.03497$\ast 10^{-5}$ \\ 
$10^{-2}$ & $-0.5$ & $4.79029\ast 10^{-9}$ & 5.80948$\ast 10^{-7}$ & 1.87121$%
\ast 10^{-7}$ & 9.07732$\ast 10^{-6}$ \\ 
$10^{-2}$ & $-0.05$ & $3.3532\ast 10^{-9}$ & 4.86056$\ast 10^{-7}$ & 1.30985$%
\ast 10^{-7}$ & 7.59463$\ast 10^{-6}$ \\ 
$10^{-2}$ & $1$ & $8.21694\ast 10^{-16}$ & 2.4477$\ast 10^{-10}$ & 1.74252$%
\ast 10^{-14}$ & 2.81794$\ast 10^{-9}$ \\ \hline
$10^{-1}$ & $-0.95$ & $6.22738\ast 10^{-8}$ & 2.09464$\ast 10^{-6}$ & 2.43257%
$\ast 10^{-6}$ & 3.27287$\ast 10^{-5}$ \\ 
$10^{-1}$ & $-0.5$ & $4.79029\ast 10^{-8}$ & 1.83712$\ast 10^{-6}$ & 1.87121$%
\ast 10^{-6}$ & 2.87050$\ast 10^{-5}$ \\ 
$10^{-1}$ & $-0.05$ & $3.3532\ast 10^{-8}$ & 1.53705$\ast 10^{-6}$ & 1.30985$%
\ast 10^{-6}$ & 2.40163$\ast 10^{-5}$ \\ 
$10^{-1}$ & $1$ & $1.77029\ast 10^{-14}$ & 1.13612$\ast 10^{-9}$ & 3.75414$%
\ast 10^{-13}$ & 1.30797$\ast 10^{-8}$ \\ \hline
$10^{0}$ & $-0.95$ & $6.22738\ast 10^{-7}$ & 6.62383$\ast 10^{-6}$ & 2.43257$%
\ast 10^{-5}$ & 1.03497$\ast 10^{-4}$ \\ 
$10^{0}$ & $-0.5$ & $4.79029\ast 10^{-7}$ & 5.80948$\ast 10^{-6}$ & 1.87121$%
\ast 10^{-5}$ & 9.07732$\ast 10^{-5}$ \\ 
$10^{0}$ & $-0.05$ & $3.3532\ast 10^{-7}$ & 4.86056$\ast 10^{-6}$ & 1.30985$%
\ast 10^{-5}$ & 7.59463$\ast 10^{-5}$ \\ 
$10^{0}$ & $1$ & $3.81397\ast 10^{-13}$ & 5.27341$\ast 10^{-9}$ & 8.08805$%
\ast 10^{-12}$ & 6.07107$\ast 10^{-8}$ \\ \hline
$10^{1}$ & $-0.95$ & $6.22738\ast 10^{-6}$ & 2.09464$\ast 10^{-5}$ & 2.43257$%
\ast 10^{-4}$ & 3.27287$\ast 10^{-4}$ \\ 
$10^{1}$ & $-0.5$ & $4.79029\ast 10^{-6}$ & 1.83712$\ast 10^{-5}$ & 1.87121$%
\ast 10^{-4}$ & 2.87050$\ast 10^{-4}$ \\ 
$10^{1}$ & $-0.05$ & $3.3532\ast 10^{-6}$ & 1.53705$\ast 10^{-5}$ & 1.30985$%
\ast 10^{-4}$ & 2.40163$\ast 10^{-4}$ \\ 
$10^{1}$ & $1$ & $8.21694\ast 10^{-12}$ & 2.4477$\ast 10^{-8}$ & 1.74252$%
\ast 10^{-10}$ & 2.81794$\ast 10^{-7}$ \\ \hline
$10^{2}$ & $-0.95$ & $6.22738\ast 10^{-5}$ & 6.62383$\ast 10^{-5}$ & 2.43257$%
\ast 10^{-3}$ & 1.03497$\ast 10^{-3}$ \\ 
$10^{2}$ & $-0.5$ & $4.79029\ast 10^{-5}$ & 5.80948$\ast 10^{-5}$ & 1.87121$%
\ast 10^{-3}$ & 9.07732$\ast 10^{-4}$ \\ 
$10^{2}$ & $-0.05$ & $3.3532\ast 10^{-5}$ & 4.86056$\ast 10^{-5}$ & 1.30985$%
\ast 10^{-3}$ & 7.59463$\ast 10^{-4}$ \\ 
$10^{2}$ & $1$ & $1.77029\ast 10^{-10}$ & 1.13612$\ast 10^{-7}$ & 3.75414$%
\ast 10^{-9}$ & 1.30797$\ast 10^{-6}$ \\ \hline
$10^{3}$ & $-0.95$ & $6.22738\ast 10^{-4}$ & 2.09464$\ast 10^{-4}$ & 2.43257$%
\ast 10^{-2}$ & 3.27287$\ast 10^{-3}$ \\ 
$10^{3}$ & $-0.5$ & $4.79029\ast 10^{-4}$ & 1.83712$\ast 10^{-4}$ & 1.87121$%
\ast 10^{-2}$ & 2.87050$\ast 10^{-3}$ \\ 
$10^{3}$ & $-0.05$ & $3.3532\ast 10^{-4}$ & 1.53705$\ast 10^{-4}$ & 1.30985$%
\ast 10^{-2}$ & 2.40163$\ast 10^{-3}$ \\ 
$10^{3}$ & $1$ & $3.81397\ast 10^{-9}$ & 5.27341$\ast 10^{-7}$ & 8.08805$%
\ast 10^{-8}$ & 6.07107$\ast 10^{-6}$ \\ \hline
$10^{4}$ & $-0.95$ & $6.22738\ast 10^{-3}$ & 6.62383$\ast 10^{-4}$ & 2.43257$%
\ast 10^{-1}$ & 1.03497$\ast 10^{-2}$ \\ 
$10^{4}$ & $-0.5$ & $4.79029\ast 10^{-3}$ & 5.80948$\ast 10^{-4}$ & 1.87121$%
\ast 10^{-1}$ & 9.07732$\ast 10^{-3}$ \\ 
$10^{4}$ & $-0.05$ & $3.3532\ast 10^{-3}$ & 4.86056$\ast 10^{-4}$ & 1.30985$%
\ast 10^{-1}$ & 7.59463$\ast 10^{-3}$ \\ 
$10^{4}$ & $1$ & $8.21694\ast 10^{-8}$ & 2.4477$\ast 10^{-6}$ & 1.74252$\ast
10^{-6}$ & 2.81794$\ast 10^{-5}$ \\ \hline
$10^{5}$ & $-0.95$ & $6.22738\ast 10^{-2}$ & 2.09464$\ast 10^{-3}$ & 2.43257$%
\ast 10^{0}$ & 3.27287$\ast 10^{-2}$ \\ 
$10^{5}$ & $-0.5$ & $4.79029\ast 10^{-2}$ & 1.83712$\ast 10^{-3}$ & 1.87121$%
\ast 10^{0}$ & 2.87050$\ast 10^{-2}$ \\ 
$10^{5}$ & $-0.05$ & $3.3532\ast 10^{-2}$ & 1.53705$\ast 10^{-3}$ & 1.30985$%
\ast 10^{0}$ & 2.40163$\ast 10^{-2}$ \\ 
$10^{5}$ & $1$ & $1.77029\ast 10^{-6}$ & 1.13612$\ast 10^{-5}$ & 3.75414$%
\ast 10^{-5}$ & 1.30797$\ast 10^{-4}$ \\ \hline
\end{tabular}
\end{table}

Although the numerical integration uses the second-order Einstein radius obtained from the cubic equation (29), the tabulated values are very close to the leading-order scaling. This is because the relevant expansion parameter is of order $r_{0}/R_{E}$, which is very small for the Galactic distances considered here. For the massive LPR branch and for the parameter ranges used in Tables 1-3, the difference between the second order $R_{E}^{(2)}$ and
\begin{equation}
  R_{E}^{(1)}=\left[2 (1 - \gamma) r_{0} \frac{D_{L} D_{LS}}{D_{S}}\right]^{1/2}
\end{equation}
is below the precision shown in the tables. For example, we show the numerical difference of the second order $R_{E}^{(2)}$ between the first order $R_{E}^{(1)}$ for massive LPR wormhole with $r_0=1$ km:
$$
R_{E}^{(2)}\big|_{\gamma = -0.95} - R_{E}^{(1)}\big|_{\gamma = -0.95} = 0.65\,{\rm km}, \; R_{E}^{(2)}\big|_{\gamma = -0.5} - R_{E}^{(1)}\big|_{\gamma = -0.5} = 0.47\,{\rm km},
\; R_{E}^{(2)}\big|_{\gamma = -0.05} - R_{E}^{(1)}\big|_{\gamma = -0.05} = 0.30 \,{\rm km}.
$$
This correction is many orders of magnitude smaller than the tabulated Einstein radii. Therefore, the table entries appear to follow the leading-order scaling even though the second-order expression was used in the numerical calculation.

%%%%%%%%%%%%%%%%%%%%%%%%%%%%%%%%%%%%%%%%%
\section{Conclusions}
\label{sec5}
%%%%%%%%%%%%%%%%%%%%%%%%%%%%%%%%%%%%%%%%%
Galactic microlensing by wormholes is an active area of research that can help to distinguish between compact astrophysical objects such as black holes and wormholes \cite{Lukmanova:2016b, Gao:2024, Tsukamoto:2018, Akhtaryanova:2024b}. The present article considers LPR phantom wormhole having a bounded mass function and the LPR parameter in the range $-1<\gamma<0$, which is related to the radial equation of state parameter by $\omega = 1/\gamma$. Our objective was to compare the microlensing profiles of phantom wormhole with the profiles of Schwarzschild black hole. We assumed the phantom wormhole to be the halo objects lensing the source stars belonging to the Galactic Bulge and to the Large Magellanic Cloud with the observer located in our galaxy. The distances involved between the observer, the lens and the source stars suggest that the concept of microlensing is well justified since the Einstein angle is too tiny, $\theta _{E}\approx 10^{-5\prime \prime }$ to $10^{-4\prime \prime }$. This fact allowed\ us to study the effect of the LPR parameter on Paczy\'{n}ski light curves and probabilistic features such as optical depth and event rate based on the hypothesis that the wormhole lens could be bound to our Galaxy.

Qualitative generic effects of LPR parameter $\gamma$ on microlensing observables are as follows:

1. The bounded LPR wormhole has ADM mass $M_{\rm ADM}=(1-\gamma)r_0/2$. The formal limit $\gamma=0$ does not give the Schwarzschild metric. It gives $A(r)=\exp(-r_0/r)$, so only the leading weak-field deflection agrees with a Schwarzschild lens of the same ADM mass, while higher-order deflection terms are different.

2. In the massive phantom branch $-1<\gamma<0$, increasing $\gamma$ decreases $M_{\rm ADM}$, the leading Einstein radius, the angular Einstein radius, and the Einstein ring crossing time.

3. The massive LPR light curves are Paczy\'{n}ski-like in the point-source approximation because the leading deflection term is proportional to $1/r$. The massless comparison case $\gamma=1$ is qualitatively different: the $1/r$ term vanishes, the leading deflection is proportional to $1/r^{2}$, and gutters can appear.

4. The Schwarzschild curves shown in the figures are reference curves for a black hole lens with the same leading ADM mass scale. They are not obtained by setting $\gamma=0$ in the LPR metric.

5. The principal robust result of the present analysis is qualitative. Massive LPR wormholes in the phantom branch behave as positive-mass compact lenses at leading order and produce Paczyński-like point-source curves. The massless comparison case has a different leading deflection order and may exhibit point-source demagnification troughs, although finite-source averaging can substantially modify these features. The optical-depth and event-rate estimates show parameter scalings within a simplified population model and are not survey forecasts. Finite-source ray tracing and realistic Galactic population and survey modeling are required before quantitative observational constraints can be derived.

\end{document}